\documentclass[aps,preprint,showpacs,preprintnumbers,amsmath,amssymb]{revtex4}
\usepackage{amsmath,mathrsfs,amsbsy,color,graphicx,bm,amsthm,amsfonts}
\usepackage{units}
\usepackage{bbm}
\usepackage{times}
\usepackage{dcolumn}
\usepackage{mathrsfs}
\usepackage{amsmath,amssymb,epsfig}
\begin{document}

\title{Topology-dependent relativistic degradation of multipartite entanglement}
\author{ Shu-Min Wu$^1$\footnote{Email: smwu@lnnu.edu.cn},  Si-Han Shang$^1$, Xu Han$^1$, Hui-Chen Yang$^1$, Qianqian Liu\footnote{Email:  qianqianliu@xtu.edu.cn  (corresponding author)}$^2$ }
\affiliation{$^1$  Department of Physics, Liaoning Normal University, Dalian 116029, China \\
$^2$ Department of Physics, Xiangtan University, Xiangtan 411105, China
}


\begin{abstract}
The influence of relativistic motion on quantum entanglement is commonly attributed to acceleration-induced thermal noise. Here we show that, for asymmetric multipartite states, the topology of entanglement can become equally important. Considering a three-qubit Star state in the Unruh-DeWitt detector framework, we compare two inequivalent acceleration configurations in which either the central or a peripheral qubit undergoes uniform acceleration. We demonstrate that these physically equivalent accelerations lead to qualitatively different entanglement dynamics: acceleration of a peripheral qubit induces a revival of one-tangle that is absent when the central qubit accelerates, whereas genuine tripartite entanglement decays monotonically but with markedly different robustness. Our results uncover a topology-dependent mechanism for relativistic entanglement degradation, showing that the response of multipartite quantum correlations is determined jointly by Unruh thermalization and the structural role of the accelerated subsystem. This work identifies asymmetric quantum networks as a distinct platform for controlling relativistic quantum resources.
\end{abstract}

\vspace*{0.5cm}
\maketitle
\section{Introduction}
Quantum entanglement, as one of the most fundamental nonclassical resources in quantum theory, plays a central role in quantum information science and provides essential advantages for quantum communication, quantum computation, quantum sensing, and quantum networks \cite{Q1,Q12,Q13,Q14,Q15,Q16,Q17,Q19,Q20,Q21}. Unlike classical correlations, entanglement reflects the intrinsic quantum correlations among subsystems and enables information-processing tasks that cannot be achieved by classical means. With the rapid development of multipartite quantum systems, understanding the generation, distribution, and preservation of multipartite entanglement has become a crucial research topic \cite{Q36,Q38}. Among various classes of multipartite entangled states, the Star state has attracted considerable attention due to its distinctive network-like structure, in which a central qubit is maximally correlated with several peripheral qubits while the outer nodes are not directly entangled with each other \cite{Q38,Q40}. This unique configuration makes the Star state an important resource for studying quantum networks, distributed quantum information processing, and entanglement sharing \cite{Q42,Q44,Q46}. Compared with conventional multipartite entangled states, the Star state exhibits characteristic entanglement properties, such as asymmetric entanglement distribution, strong central-node correlations, and robustness against certain types of local perturbations. These features make it an ideal candidate for exploring the behavior of multipartite entanglement under realistic physical environments.

By bridging quantum information theory, quantum field theory, and general relativity, relativistic quantum information has emerged as a powerful framework for exploring the behavior of quantum resources in noninertial motion and curved spacetime \cite{SDF1,SDF2,SDF3,SDF4,SDF5,SDF6,SDF7,SDF8,SDF9,SDF10,SDF11,SDF12,SDF13,SDF14,SDF15,SDF16,SDF17,SDF18,SDF19,SDF20,SDF21,SDF22,SDF23,SDF24,SDF25,SDF26,SDF27,SDF28,SDF29,SDF30,SDF31,SDF32,SDF33,SDF34,SDF35,SDF36,SDF37,SDF38,SDF39,SDF40,SDF41,SDF42,SDF43}. Previous studies have established that relativistic effects, such as the Unruh and Hawking effects, generally degrade multipartite entanglement \cite{SDF45,SDF46,SDF47,SDF48,SDF49,SDF50}. However, an important question remains largely unexplored: Is relativistic entanglement degradation determined solely by the strength of the relativistic effect, or can it also depend on the topology of multipartite quantum correlations?
Existing investigations have almost exclusively focused on permutation-symmetric states, such as the GHZ and W states \cite{SDF45,SDF46,SDF47,SDF48,SDF49,SDF50,SDF51,SDF52,SDF53,SDF54,SDF55}. Because all subsystems in these states are physically equivalent, accelerating different qubits leads to identical entanglement dynamics, making it impossible to distinguish whether the response is governed purely by relativistic thermalization or also by the structural role of the accelerated subsystem. Consequently, the influence of entanglement topology on relativistic quantum resources has remained hidden. Asymmetric multipartite states provide the natural setting for addressing this issue. Among them, the Star state possesses a distinctive topology consisting of one central qubit connected to multiple peripheral qubits, whose nonequivalent structural roles offer a direct means of isolating subsystem-dependent relativistic effects. Accelerating different nodes therefore enables one to disentangle the contributions of relativistic thermal noise and entanglement topology, providing new insight into how multipartite quantum resources are redistributed in relativistic environments.

To address this question in an operationally realistic setting, we employ the Unruh-DeWitt detector model, which describes localized interactions between finite-dimensional quantum systems and quantum fields without relying on the idealized assumption of globally excited field modes \cite{SDF56,SDF57,SDF58}. Within this framework, we investigate the relativistic dynamics of tripartite entanglement in an asymmetric Star state by considering two physically equivalent acceleration configurations, in which either the central or a peripheral detector undergoes uniform acceleration. This setup enables us to isolate the influence of the structural role of the accelerated subsystem while keeping the relativistic effect itself unchanged. We show that identical accelerations acting on different nodes produce qualitatively distinct entanglement dynamics, demonstrating that relativistic degradation is governed jointly by Unruh-induced thermalization and the topology of multipartite entanglement, rather than by acceleration alone. These findings establish subsystem-dependent relativistic dynamics as a fundamental feature of asymmetric multipartite quantum systems and suggest that engineering entanglement topology may provide a viable route for controlling quantum resources in relativistic environments.

The remainder of this paper is organized as follows. Section II introduces the relativistic evolution of the asymmetric Star state within the Unruh-DeWitt detector framework, where one detector undergoes uniform acceleration. In Section III, we investigate the effects of relativistic motion on multipartite quantum correlations by analyzing the behavior of the one-tangle and genuine tripartite entanglement. Section IV concludes the paper with a summary of the main findings and their physical implications.

\section{Evolution of the Star state under relativistic motion}
\begin{figure}
\centering
\includegraphics[height=2.3in,width=5.6in]{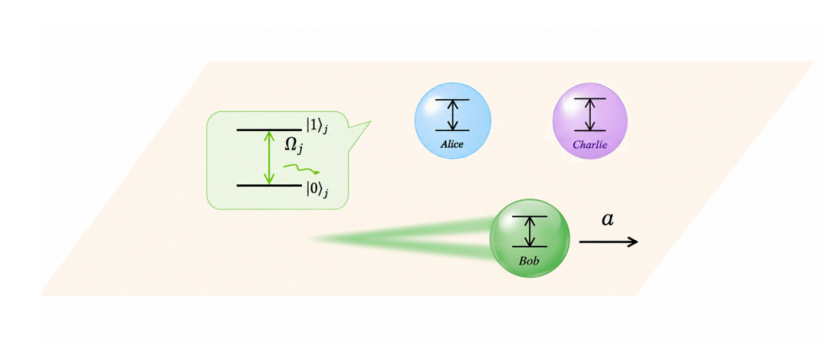}
\caption{Schematic diagram of the tripartite Unruh-DeWitt detector system. Three detectors, labeled Alice, Bob, and Charlie, are initially prepared in a three-qubit Star state. Bob's detector undergoes uniform proper acceleration $a$, while Alice's and Charlie's detectors remain inertial. Each detector is modeled as a two-level system with an energy gap $\Omega$.}
\label{F1}
\end{figure}

In this paper, we consider a tripartite system consisting of three observers, Alice ($A$), Bob ($B$), and Charlie ($C$), each equipped with an Unruh-DeWitt detector modeled as a noninteracting two-level atom. As the initial resource state, we choose the three-qubit Star state,
\begin{eqnarray}\label{w1}
|\text{Star}\rangle=\frac{1}{2}(|000\rangle+|100\rangle+|101\rangle+|111\rangle),
\end{eqnarray}
where $|0\rangle$ and $|1\rangle$ denote the ground and excited states of each detector, respectively \cite{Q38}. Unlike permutation-symmetric multipartite states, the Star state possesses an intrinsically asymmetric entanglement topology composed of two peripheral qubits and one central qubit. Throughout this work, Alice and Bob are regarded as the peripheral qubits, while Charlie serves as the central qubit. This asymmetry is reflected in the distribution of quantum correlations: tracing out the central qubit leaves the reduced state of the two peripheral qubits separable, whereas tracing out either peripheral qubit preserves bipartite entanglement between the remaining qubits. Consequently, the Star state provides an ideal platform for investigating how the structural role of the accelerated subsystem influences relativistic entanglement dynamics. We consider two physically equivalent acceleration configurations. In the first scenario, Bob's detector undergoes uniform acceleration while Alice and Charlie remain inertial in Fig.\ref{F1}; in the second, Charlie is uniformly accelerated with Alice and Bob remaining at rest. Since the derivations for the two cases follow the same procedure, we present the former in the main text, while the latter is summarized in Appendix A.
We assume that Bob's detector undergoes uniform proper acceleration $a$ for a finite proper-time interval $\Delta$ (see Fig.\ref{F1}), while the detectors of Alice and Charlie remain inertial and switched off. Consequently, only Bob's detector actively couples to the external field during this interval. The worldline of Bob's detector is parametrized by
\begin{eqnarray}\label{w2}
t(\tau)=a^{-1}\sinh a\tau,  \quad x(\tau)=a^{-1}\cosh a\tau,  \quad  y(\tau)=z(\tau)=0,
\end{eqnarray}
where $a$ represents Bob's proper acceleration and $\tau$ denotes the detector's proper time. Throughout this paper, we adopt the natural units $c=\hbar=k_{B}=1$ for simplicity \cite{SDF58}. The initial state of the composite detector-field system is given by
\begin{eqnarray}\label{ww3}
|\text{Star}_{-\infty}^{ABC\phi}\rangle=|\text{Star}\rangle\otimes|0_{M}\rangle,
\end{eqnarray}
where $|0_{M}\rangle$ represents the Minkowski vacuum state of the external scalar field.

The total Hamiltonian of the composite system is written as
\begin{eqnarray}\label{w4}
H_{ABC\phi}=H_{A}+H_{B}+H_{C}+H_{KG}+H_{\text{int}}^{B\phi}.
\end{eqnarray}
Here, $H_{KG}$ represents the Hamiltonian of the massless scalar field, $H_{\rm{int}}^{B\phi}$ describes the coupling between Bob's detector and the field, while $H_P = \Omega P^\dagger P$ ($P \in \{A, B, C\}$) denotes the free internal Hamiltonian of each detector, with $\Omega$ being the energy gap between the ground state $|0\rangle$ and the excited state $|1\rangle$ \cite{SDF59}. The corresponding ladder operators satisfy $P^\dagger |0\rangle = |1\rangle$, $P |1\rangle = |0\rangle$, and $P |0\rangle = P^\dagger |1\rangle = 0$. The interaction between Bob's accelerated detector and the massless scalar field $\phi(x)$ is governed by the localized interaction Hamiltonian
\begin{eqnarray}\label{w5}
H_{\text{int}}^{B\phi}(t)=\epsilon(t)\int_{\Sigma_{t}}\text{d}^{3} \boldsymbol{x}\sqrt{-g}\phi(x)[\chi(\boldsymbol{x})B+\bar{\chi}(\boldsymbol{x})B^{\dagger}],
\end{eqnarray}
where $g \equiv \det(g_{ab})$ and $g_{ab}$ is the Minkowski spacetime metric. The function $\chi(\boldsymbol{x}) = (\kappa\sqrt{2\pi})^{-3}\text{exp}(-\boldsymbol{x}^{2}/2\kappa^{2})$ is a Gaussian coupling function with a constant characteristic width $\kappa$, signifying that the detector interacts exclusively with its neighboring field.

In the weak-coupling regime, the final state $|\text{Star}^{ABC\phi}_\infty\rangle$ of the total system can be calculated using first-order perturbation theory with respect to the coupling constant $\epsilon$, yielding
\begin{eqnarray}\label{w6}
|\text{Star}^{ABC\phi}_\infty\rangle=\{I+a^{\dagger}_{RI}(\lambda)B-a_{RI}
(\overline{\lambda})B^{\dagger}\}|\text{Star}_{-\infty}^{ABC\phi}\rangle,
\end{eqnarray}
where the operators $a^{\dagger}_{RI}$ and $a_{RI}$ denote the creation and annihilation operators for the $\lambda$ modes in Rindler region $I$, respectively.
The function $f$ is defined as $f\equiv\epsilon(t)e^{-i\Omega t}\chi(\boldsymbol{x})$.
The operator $K$ establishes the correspondence between the positive-frequency part of the solutions of the Klein-Gordon equation $\nabla_{a}\nabla^{a}\phi(x)=0$ and the timelike isometry. The term $Ef$ can be expressed as
\begin{eqnarray}\label{qq2}
Ef=\int \text{d}^{4}x^{\prime}\sqrt{-g(x^{\prime})}[G^{\rm{adv}}(x,x^{\prime})-G^{\rm{ret}}(x,x^{\prime})]f(x^{\prime}),
\end{eqnarray}
where $E$ is the difference between the advanced Green's function $G^{\rm{adv}}$ and the retarded Green's function $G^{\rm{ret}}$.

Substituting the initial state $|\text{Star}_{-\infty}^{ABC\phi}\rangle$ from Eq.(\ref{ww3}) into Eq.(\ref{w6}) yields the final state of the total system in terms of the Rindler operators:
\begin{eqnarray}\label{w7}
|\text{Star}^{ABC\phi}_\infty\rangle&=&|\text{Star}_{-\infty}^{ABC\phi}\rangle+\frac{1}{2}
\Big[|101\rangle\otimes(a^{\dagger}_{RI}(\lambda)|0_{M}\rangle)\\
&+&(|010\rangle+|110\rangle+|111\rangle)\otimes(a_{RI}
(\overline{\lambda})|0_{M}\rangle)\Big]\notag.
\end{eqnarray}
Here, the Rindler operators $a_{RI}(\overline{\lambda})$ and $a^{\dagger}_{RI}(\lambda)$ are defined in Rindler region $I$, and $|0_{M}\rangle$ is the vacuum state in the Minkowski spacetime. The Bogoliubov transformations relating the Rindler operators to the Minkowski modes are given by
\begin{eqnarray}\label{w8}
a_{RI}(\overline{\lambda})=\frac{a_{M}(\overline{F_{1\Omega}})+\text{e}^{-\pi\Omega/a}
a^{\dagger}_{M}(F_{2\Omega})}{(1-\text{e}^{-2\pi\Omega/a})^{1/2}},
\end{eqnarray}

\begin{eqnarray}\label{w9}
a^{\dagger}_{RI}(\lambda)=\frac{a^{\dagger}_{M}({F_{1\Omega}})+
\text{e}^{-\pi\Omega/a}a_{M}(\overline{F_{2\Omega}})}{(1-\text{e}^{-2\pi\Omega/a})^{1/2}},
 \end{eqnarray}
where $F_{1\Omega}=\frac{\lambda+\text{e}^{-\pi\Omega/a}\lambda\circ\omega}
{(1-\text{e}^{-2\pi\Omega/a})^{1/2}}$ and $F_{2\Omega}=\frac{{\overline{\lambda\circ\omega}+\text{e}^{-\pi\Omega/a}\overline{\lambda}}}
{(1-\text{e}^{-2\pi\Omega/a})^{1/2}}$.
Here, $\omega(t,x,y,z)=(-t,-x,y,z)$ represents the wedge-reflection isometry mapping the mode $\lambda$ from Rindler region $I$ to $\lambda\circ\omega$ in region $II$, where the symbol $\circ$ denotes map composition.

Utilizing the Bogoliubov transformations provided in Eqs.(\ref{w8}) and (\ref{w9}), Eq.(\ref{w7}) can be reformulated as
\begin{equation}
\label{w10}
\begin{aligned}
|\mathrm{Star}^{ABC\phi}_\infty\rangle = &|\mathrm{Star}_{-\infty}^{ABC\phi}\rangle + \frac{1}{2}\nu\bigg[\frac{|101\rangle\otimes|1_{\tilde{F}_{1\Omega}}\rangle}{(1-\text{e}^{-2\pi\Omega/a})^{1/2}} \\
&+ \text{e}^{-\pi\Omega/a}\frac{(|010\rangle + |110\rangle + |111\rangle)\otimes |1_{\tilde{F}_{2\Omega}}\rangle}{(1-\mathrm{e}^{-2\pi\Omega/a})^{1/2}}\bigg],
\end{aligned}
\end{equation}
where $\tilde{F}_{i\Omega}=F_{i\Omega}/\nu$.
Since we are primarily interested in the state of the detectors after Bob's acceleration, we trace out the degrees of freedom associated with the external scalar field. The resulting reduced density matrix of the tripartite system is
\begin{eqnarray}\label{w18}
\rho_{ABC}=\|\text{Star}^{ABC\phi}_{\infty}\|^{-2}\text{tr}_{\phi}|\text{Star}^{ABC\phi}_{\infty}\rangle
\langle\text{Star}^{ABC\phi}_{\infty}|,
\end{eqnarray}
where the normalization factor $\|\text{Star}^{ABC\phi}_{\infty}\|^{2}=1+\frac{\nu^{2}(1+3\text{e}^{-2\pi\Omega/a})}
{4(1-\text{e}^{-2\pi\Omega/a})}$ ensures $\text{tr}(\rho_{ABC})=1$. Explicitly, the final detector state takes the form
\begin{eqnarray}\label{w20}
\rho_{ABC}=
 \left(\!\!\begin{array}{cccccccc}
Q_{0}&0&0&0&Q_{0}&Q_{0}&0&Q_{0}\\
0&0&0&0&0&0&0&0\\
0&0&Q_{2}&0&0&0&Q_{2}&Q_{2}\\
0&0&0&0&0&0&0&0\\
Q_{0}&0&0&0&Q_{0}&Q_{0}&0&Q_{0}\\
Q_{0}&0&0&0&Q_{0}&Q_{0}+Q_{1}&0&Q_{0}\\
0&0&Q_{2}&0&0&0&Q_{2}&Q_{2}\\
Q_{0}&0&Q_{2}&0&Q_{0}&Q_{0}&Q_{2}&Q_{0}+Q_{2}\\
\end{array}\!\!\right),
\end{eqnarray}
where the coefficients $Q_{0}$, $Q_{1}$, and $Q_{2}$ are given by
\begin{eqnarray}\label{w21}
Q_{0}=\frac{1-\text{e}^{-2\pi\Omega/a}}{\nu^{2}(1+3\text{e}^{-2\pi\Omega/a})+4(1-\text{e}^{-2\pi\Omega/a})}\notag,
\end{eqnarray}
\begin{eqnarray}\label{w22}
Q_{1}=\frac{\nu^{2}}{\nu^{2}(1+3\text{e}^{-2\pi\Omega/a})+4(1-\text{e}^{-2\pi\Omega/a})}\notag,
\end{eqnarray}
\begin{eqnarray}\label{w233}
Q_{2}=\frac{\nu^{2}\text{e}^{-2\pi\Omega/a}}{\nu^{2}(1+3\text{e}^{-2\pi\Omega/a})+4(1-\text{e}^{-2\pi\Omega/a})}\notag.
\end{eqnarray}
Here, the acceleration parameter $q\equiv\text{e}^{-2\pi\Omega/a}$ and the effective coupling strength $\nu^{2}\equiv\|\lambda\|^{2}=\frac{\epsilon^{2}\Omega\Delta}{2\pi}\text{e}^{-\Omega^{2}\kappa^{2}}$.
These expressions are valid under the conditions $\epsilon\ll\Omega^{-1}\ll\Delta$ and $\epsilon$ varies slowly with time compared to the frequency $\Omega$.
To ensure the validity of the first-order perturbative approach, the effective coupling must satisfy $\nu^{2}\ll1$.
Additionally, the parameter $q$ is a monotonic function of the acceleration $a$; the limit $q\rightarrow0$ corresponds to the inertial case (zero acceleration), whereas $q\rightarrow1$ represents the asymptotic limit of infinite acceleration.

\section{Quantum  entanglement of the Star state in relativistic motion}
\subsection{ One-tangle under relativistic motion}
Negativity quantifies entanglement by assessing how much the partially transposed density matrix of a system contains negative eigenvalues \cite{SDF60,SDF61}. In a tripartite system, the entanglement between a single subsystem and the remaining two subsystems is referred to as the one-tangle, which is defined as
\begin{eqnarray}\label{ww26}
N_{\alpha(\beta\gamma)}=\parallel\rho^{T_{\alpha}}_{\alpha\beta\gamma}\parallel-1.
\end{eqnarray}
Here, $T_{\alpha}$ represents the partial transpose of $\rho_{\alpha\beta\gamma}$ with respect to subsystem $\alpha$.
Noting that $\|\alpha\| - 1$ equals twice the sum of the absolute values of the negative eigenvalues, the one-tangle can equivalently be written as
\begin{eqnarray}\label{ww266}
N_{\alpha(\beta\gamma)}=2\sum^{n}_{i=1}|\lambda^{(-)}_{\alpha(\beta\gamma)}|^{i},
\end{eqnarray}
where $|\lambda^{(-)}_{\alpha(\beta\gamma)}|^{i}$  are the negative eigenvalues of the partially transposed matrix.

We analyze the negativity of the tripartite Star state described by the density operator $\rho_{ABC}$ by taking the partial transpose with respect to subsystem $A$, which yields the following expression

\begin{eqnarray}\label{w20}
\rho^{T_{A}}_{ABC}=
 \left(\!\!\begin{array}{cccccccc}
Q_{0}&0&0&0&Q_{0}&0&0&0\\
0&0&0&0&Q_{0}&0&0&0\\
0&0&Q_{2}&0&0&0&Q_{2}&0\\
0&0&0&0&Q_{0}&0&Q_{2}&0\\
Q_{0}&Q_{0}&0&Q_{0}&Q_{0}&Q_{0}&0&Q_{0}\\
0&0&0&0&Q_{0}&Q_{0}+Q_{1}&0&Q_{0}\\
0&0&Q_{2}&Q_{2}&0&0&Q_{2}&Q_{2}\\
0&0&0&0&Q_{0}&Q_{0}&Q_{2}&Q_{0}+Q_{2}\\
\end{array}\!\!\right).
\end{eqnarray}
Similarly, by taking the transpose with respect to the $B$ mode, we obtain
\begin{eqnarray}\label{w20}
\rho^{T_{B}}_{ABC}=
 \left(\!\!\begin{array}{cccccccc}
Q_{0}&0&0&0&Q_{0}&Q_{0}&0&0\\
0&0&0&0&0&0&0&0\\
0&0&Q_{2}&0&0&Q_{0}&Q_{2}&Q_{0}\\
0&0&0&0&0&0&0&0\\
Q_{0}&0&0&0&Q_{0}&Q_{0}&0&0\\
Q_{0}&0&Q_{0}&0&Q_{0}&Q_{0}+Q_{1}&Q_{0}&Q_{0}\\
0&0&Q_{2}&0&0&Q_{0}&Q_{2}&Q_{2}\\
0&0&Q_{2}&0&0&Q_{0}&Q_{2}&Q_{0}+Q_{2}\\
\end{array}\!\!\right).
\end{eqnarray}
Alternatively, performing the transpose with respect to the $C$ mode, we obtain
\begin{eqnarray}\label{w20}
\rho^{T_{C}}_{ABC}=
 \left(\!\!\begin{array}{cccccccc}
Q_{0}&0&0&0&Q_{0}&0&0&0\\
0&0&0&0&Q_{0}&0&Q_{0}&0\\
0&0&Q_{2}&0&0&0&Q_{2}&0\\
0&0&0&0&0&0&Q_{2}&0\\
Q_{0}&Q_{0}&0&0&Q_{0}&Q_{0}&0&0\\
0&0&0&0&Q_{0}&Q_{0}+Q_{1}&Q_{0}&Q_{0}\\
0&Q_{0}&Q_{2}&Q_{2}&0&Q_{0}&Q_{2}&Q_{2}\\
0&0&0&0&0&Q_{0}&Q_{2}&Q_{0}+Q_{2}\\
\end{array}\!\!\right).
\end{eqnarray}
Owing to the intricate form of the negativity calculation for the Star state, we omit the explicit results here.

\begin{figure}[htbp]
\centering

\begin{minipage}{2.1in}
\centering
(a)
\end{minipage}
\begin{minipage}{2.1in}
\centering
(b)
\end{minipage}
\begin{minipage}{2.1in}
\centering
(c)
\end{minipage}

\vspace{0.05cm}

\includegraphics[height=1.8in,width=2.1in]{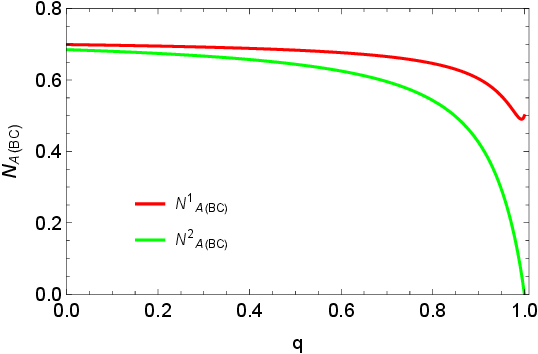}
\includegraphics[height=1.8in,width=2.1in]{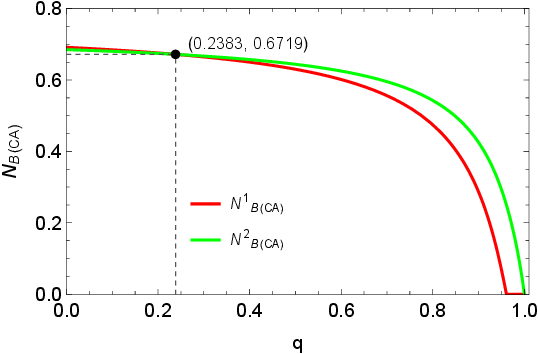}
\includegraphics[height=1.79in,width=2.1in]{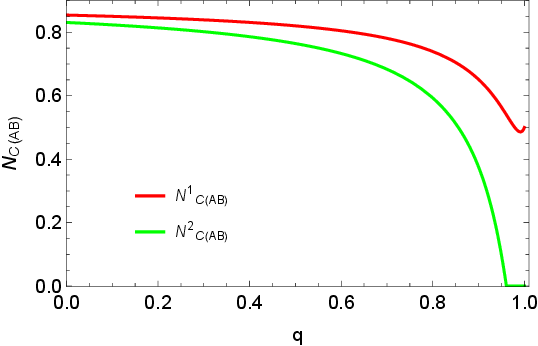}

\caption{The  one-tangle $N^{1/2}_{A(BC)}$, $N^{1/2}_{B(CA)}$, and $N^{1/2}_{C(AB)}$ as functions of the acceleration $q$ at fixed coupling parameter $\nu^{2} = 0.04$, where $N^1$ and $N^2$ denote the cases in which Bob's and Charlie's detectors undergo uniform acceleration, respectively.}
\label{F2}
\end{figure}

In Fig.\ref{F2}(a), we plot the one-tangle $N_{A(BC)}$ as  functions of the acceleration parameter $q$ for the two acceleration configurations. When Bob undergoes uniform acceleration, $N^1_{A(BC)}$ exhibits a non-monotonic dependence on the acceleration. Specifically, the
one-tangle initially decreases due to the combined influence of the Unruh-induced thermal noise and the redistribution of  quantum correlations, reaches a minimum at higher acceleration, and then gradually increases as $q$ approaches the infinite-acceleration limit. By contrast, when Charlie is accelerated, $N^2_{A(BC)}$ decreases monotonically with increasing acceleration and eventually vanishes in the limit $q\rightarrow1$. These distinct behaviors indicate that the evolution of the one-tangle depends not only on the strength of the relativistic effect but also on which subsystem is subjected to acceleration. Although Bob and Charlie experience the same acceleration, their different roles in the Star-state entanglement structure lead to qualitatively different responses of the bipartite entanglement.

Fig\ref{F2}(b) shows the acceleration dependence of the one-tangle $N_{B(CA)}$. For small accelerations, $N^1_{B(CA)}$ is larger than $N^2_{B(CA)}$, indicating that the entanglement associated with Bob is initially more robust when Bob rather than Charlie undergoes acceleration. As the acceleration increases, the two curves intersect at the critical value $q\approx0.2383$, beyond which $N^2_{B(CA)}$ exceeds $N^1_{B(CA)}$. Furthermore, $N^1_{B(CA)}$ eventually undergoes entanglement sudden death at sufficiently large acceleration, whereas $N^2_{B(CA)}$ remains finite over the same parameter range. This crossover demonstrates that the influence of relativistic motion on quantum entanglement is strongly dependent on the accelerated subsystem. In particular, the direct thermal decoherence acting on Bob becomes increasingly dominant at large acceleration, leading to a more pronounced suppression of the one-tangle than the indirect influence arising from Charlie's acceleration.

Finally, Fig.\ref{F2}(c) presents the behavior of the one-tangle $N_{C(AB)}$. Similar to the behavior of $N^1_{A(BC)}$ in Fig.~\ref{F2}(a), $N^1_{C(AB)}$ first decreases and then increases as the acceleration approaches the infinite-acceleration limit, exhibiting a clear revival of entanglement. In contrast, when Charlie undergoes acceleration, $N^2_{C(AB)}$ decreases monotonically and eventually undergoes sudden death without any revival. The different behaviors observed in the two acceleration configurations further confirm that the relativistic evolution of quantum entanglement is subsystem dependent. The distinct positions of Bob and Charlie within the asymmetric Star state result in different sensitivities to the Unruh effect, giving rise to markedly different entanglement dynamics. This subsystem-dependent relativistic behavior reveals the crucial role of  entanglement structures in determining the evolution of quantum resources and suggests that engineering suitable quantum network architectures may provide an effective strategy for enhancing the robustness and controllability of quantum resources in relativistic environments.

\subsection{Genuine tripartite entanglement under relativistic motion}
For a tripartite system, genuine multipartite entanglement can be characterized through the monogamy relation of quantum entanglement. The Coffman-Kundu-Wootters (CKW) inequality establishes a fundamental constraint on the distribution of entanglement among different subsystems \cite{SDF61}, stating that the entanglement shared between one subsystem and the remaining two cannot be completely concentrated into pairwise correlations. Instead, a portion of the quantum correlations is intrinsically multipartite. The CKW inequality is expressed as
\begin{equation}
[N_{\alpha(\beta\gamma)}]^2
\geq
[N_{\alpha\beta}]^2
+
[N_{\alpha\gamma}]^2,
\end{equation}
where $N_{\alpha(\beta\gamma)}$ denotes the one-tangle between subsystem $\alpha$ and the composite subsystem $(\beta\gamma)$, while $N_{\alpha\beta}$ and $N_{\alpha\gamma}$ represent the corresponding two-tangles.

The difference between the two sides of the CKW inequality, known as the residual entanglement, quantifies the quantum correlations that cannot be attributed to any pairwise entanglement and therefore provides a measure of genuine tripartite entanglement. Following this idea, we adopt the minimally residual entanglement as the measure of genuine tripartite entanglement,
\begin{equation}\label{ep6}
E_{(A|B|C)}
=
\min_{(A,B,C)}
\left\{
[N_{A(BC)}]^2
-
[N_{AB}]^2
-
[N_{AC}]^2
\right\},
\end{equation}
where the minimum is taken over all permutations of the subsystem labels $(A,B,C)$. This definition ensures permutation invariance of the entanglement measure and provides a faithful quantification of the genuine tripartite quantum correlations shared among the three subsystems.

\begin{figure}
\begin{minipage}[t]{0.5\linewidth}
\centering
\includegraphics[width=3.0in,height=5.2cm]{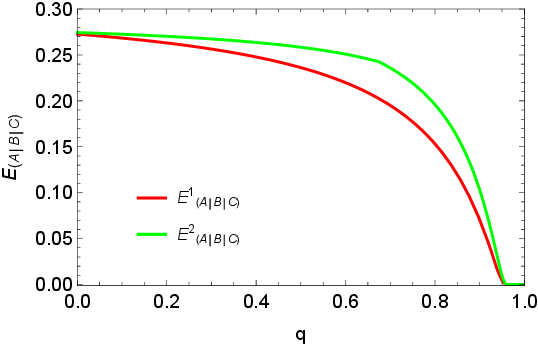}
\label{fig4}
\end{minipage}%

\caption{The genuine tripartite entanglement $E_{(A|B|C)}^{1/2}$ versus the acceleration parameter $q$. The superscripts $1$ and $2$ denote the cases in which Bob's and Charlie's detectors undergo uniform acceleration, respectively.}
\label{Fig4}
\end{figure}

As shown in Fig.\ref{Fig4}, the genuine tripartite entanglement  decreases monotonically with increasing acceleration parameter $q$ in both acceleration configurations and eventually suffers from sudden death, indicating the irreversible degradation of genuine multipartite quantum correlations by the Unruh effect. Unlike the one-tangle, no entanglement revival occurs for the genuine tripartite entanglement, suggesting that the redistribution of quantum correlations induced by relativistic motion is insufficient to compensate for the loss of genuine multipartite entanglement. Throughout the entire acceleration regime, the configuration with Charlie undergoing acceleration always retains a larger genuine tripartite entanglement than that with Bob accelerating, despite the two detectors experiencing identical relativistic motion.
This contrast demonstrates that the degradation of genuine multipartite entanglement is not determined solely by Unruh-induced thermalization, but also by the structural role of the accelerated subsystem within the entanglement topology. In the Star state, accelerating the central and peripheral nodes affects the redistribution of quantum correlations in fundamentally different ways, leading to distinct robustness of multipartite entanglement under identical relativistic conditions. These results identify entanglement topology as an independent physical ingredient governing relativistic quantum correlations and reveal that subsystem asymmetry provides a mechanism for controlling multipartite quantum resources in relativistic settings.

\section{Conclusions}
We have investigated the relativistic dynamics of multipartite entanglement in an asymmetric three-qubit Star state using the Unruh-DeWitt detector model. By comparing two physically equivalent acceleration configurations, we have demonstrated that the response of multipartite entanglement depends crucially on which subsystem experiences relativistic motion.
Specifically, while the one-tangle exhibits either revival or monotonic degradation depending on the accelerated detector, the genuine tripartite entanglement decreases monotonically in both cases but displays markedly different robustness. These distinct behaviors reveal that the evolution of multipartite entanglement is jointly determined by Unruh-induced thermalization and the topology of the underlying entanglement structure, rather than by acceleration alone.
Our results establish entanglement topology as an essential ingredient in relativistic quantum information beyond permutation-symmetric multipartite states. They also suggest that exploiting asymmetric entanglement structures may provide an effective strategy for protecting and controlling quantum resources in relativistic quantum communication, distributed quantum information processing, and other relativistic quantum technologies.

\begin{acknowledgments}
This work is supported by the National Natural Science Foundation of China (12575056),  the Natural Science Foundation of Hunan Province under Grant No.2025JJ60019, the Natural
Science Foundation of the Liaoning Scientific Committee No. 2026-MS-281, and LiaoNing Revitalization Talents Program (XLYC2503099).	
\end{acknowledgments}


\appendix
\section{Derivations for Charlie's acceleration scenario}
\begin{figure}
\centering
\includegraphics[height=2.1in,width=5.6in]{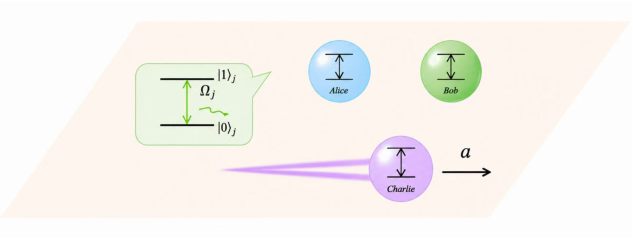}
\caption{Schematic illustration of the tripartite Unruh-DeWitt detector system, where Alice, Bob, and Charlie initially share a three-qubit Star state and only Charlie's detector undergoes uniform proper acceleration $a$.}
\label{F6}
\end{figure}

In this appendix, we consider the alternative scenario mentioned in the section II, where Charlie's detector undergoes uniform proper acceleration while the detectors of Alice and Bob remain stationary in Fig.\ref{F6}.
In the weak-coupling regime, the final state $|\widetilde{\text{Star}}^{ABC\phi}_\infty\rangle$ of the atom-field system can be calculated in first-order perturbation theory, which can be expressed as
\begin{eqnarray}\label{w66}
|\widetilde{\text{Star}}^{ABC\phi}_\infty\rangle=\{I+a^{\dagger}_{RI}(\lambda)C-a_{RI}
(\overline{\lambda})C^{\dag}\}|\text{Star}_{-\infty}^{ABC\phi}\rangle.
\end{eqnarray}

Substituting the initial state $|\text{Star}_{-\infty}^{ABC\phi}\rangle$ from Eq.(\ref{ww3}) into Eq.(\ref{w66}) yields the final state of the total system as
\begin{eqnarray}\label{AA2}
|\widetilde{\text{Star}}^{ABC\phi}_\infty\rangle&=&|{\text{Star}}_{-\infty}^{ABC\phi}\rangle+\frac{1}{2}\Big[{(|100\rangle+|110\rangle)\otimes({a^\dagger_{RI}}(\lambda)|0_{M}\rangle})\\
&+&{(|001\rangle+|101\rangle)\otimes({a_{RI}}(\overline{\lambda})|0_{M}\rangle})\Big]\notag,
\end{eqnarray}
where $a^{\dagger}_{RI}(\lambda)$ and $a_{RI}(\overline{\lambda})$ denote the creation and annihilation operators associated with Rindler region $I$, and $|0_{M}\rangle$ corresponds to the Minkowski vacuum state.
By applying the Bogoliubov transformations given in Eqs.(\ref{w8}) and (\ref{w9}), Eq.(\ref{AA2}) can be reformulated as
\begin{eqnarray}\label{w16}
|\widetilde{\text{Star}}^{ABC\phi}_\infty\rangle&=&|{\text{Star}}_{-\infty}^{ABC\phi}\rangle+\frac{1}{2}\nu\bigg[\frac{(|100\rangle+|110\rangle)\otimes|1_{\tilde{F}_{1\Omega}}\rangle}{(1-\textrm{e}^{-2\pi\Omega/a})^{1/2}}\\
&+&\textrm{e}^{-\pi\Omega/a}\frac{(|001\rangle+|101\rangle)\otimes|1_{\tilde{F}_{2\Omega}}\rangle}{(1-\textrm{e}^{-2\pi\Omega/a})^{1/2}}\bigg]\notag,
\end{eqnarray}
where $\tilde{F}_{i\Omega}=F_{i\Omega}/\nu$.

To determine the state of the detectors after their interaction with the field, we trace out the external field degrees of freedom, yielding the reduced density matrix of the system $ABC$
\begin{eqnarray}\label{p4}
\rho'_{ABC}=\|\widetilde{\text{Star}}^{ABC\phi}_{\infty}\|^{-2}\textrm{tr}_{\phi}|\widetilde{\text{Star}}^{ABC\phi}_{\infty}\rangle
\langle\widetilde{\text{Star}}^{ABC\phi}_{\infty}|,
\end{eqnarray}
where the normalization coefficient  $\|\widetilde{\text{Star}}^{ABC\phi}_{\infty}\|^{2}=1+\frac{\nu^{2}(1+\textrm{e}^{-2\pi\Omega/a})}
{2(1-\textrm{e}^{-2\pi\Omega/a})}$ ensures $\text{tr}(\rho'_{ABC})=1$. Therefore, when Charlie's detector undergoes uniform acceleration, the final state of the detectors is found to be
\begin{eqnarray}\label{w20}
\rho'_{ABC}=
 \left(\!\!\begin{array}{cccccccc}
R_{0}&0&0&0&R_{0}&R_{0}&0&R_{0}\\
0&R_{2}&0&0&0&R_{2}&0&0\\
0&0&0&0&0&0&0&0\\
0&0&0&0&0&0&0&0\\
R_{0}&0&0&0&R_{0}+R_{1}&R_{0}&R_{1}&R_{0}\\
R_{0}&R_{2}&0&0&R_{0}&R_{0}+R_{2}&0&R_{0}\\
0&0&0&0&R_{1}&0&R_{1}&0\\
R_{0}&0&0&0&R_{0}&R_{0}&0&R_{0}\\
\end{array}\!\!\right),
\end{eqnarray}
where the coefficients $R_{0}$, $R_{1}$, and $R_{2}$ are explicitly given by
$$R_{0}=\frac{1-\textrm{e}^{-2\pi\Omega/a}}{2\nu^{2}(1+\textrm{e}^{-2\pi\Omega/a})+4(1-\textrm{e}^{-2\pi\Omega/a})}\notag,$$
$$R_{1}=\frac{\nu^{2}}{2\nu^{2}(1+\textrm{e}^{-2\pi\Omega/a})+4(1-\textrm{e}^{-2\pi\Omega/a})}\notag,$$
$$R_{2}=\frac{\nu^{2}\textrm{e}^{-2\pi\Omega/a}}{2\nu^{2}(1+\textrm{e}^{-2\pi\Omega/a})+4(1-\textrm{e}^{-2\pi\Omega/a})}\notag.$$
The partial transpositions of the reduced density matrix $\rho'_{ABC}$ with respect to subsystems $A$, $B$, and $C$ are explicitly derived as
\begin{eqnarray}\label{rho_TA}
\rho'^ {T_{A}}_{ABC}=
 \left(\!\!\begin{array}{cccccccc}
R_{0}&0&0&0&R_{0}&0&0&0\\
0&R_{2}&0&0&R_{0}&R_{2}&0&0\\
0&0&0&0&0&0&0&0\\
0&0&0&0&R_{0}&0&0&0\\
R_{0}&R_{0}&0&R_{0}&R_{0}+R_{1}&R_{0}&R_{1}&R_{0}\\
0&R_{2}&0&0&R_{0}&R_{0}+R_{2}&0&R_{0}\\
0&0&0&0&R_{1}&0&R_{1}&0\\
0&0&0&0&R_{0}&R_{0}&0&R_{0}\\
\end{array}\!\!\right),
\end{eqnarray}
\begin{eqnarray}\label{rho_TB}
\rho'^ {T_{B}}_{ABC}=
 \left(\!\!\begin{array}{cccccccc}
R_{0}&0&0&0&R_{0}&R_{0}&0&0\\
0&R_{2}&0&0&0&R_{2}&0&0\\
0&0&0&0&0&R_{0}&0&0\\
0&0&0&0&0&0&0&0\\
R_{0}&0&0&0&R_{0}+R_{1}&R_{0}&R_{1}&0\\
R_{0}&R_{2}&R_{0}&0&R_{0}&R_{0}+R_{2}&R_{0}&R_{0}\\
0&0&0&0&R_{1}&R_{0}&R_{1}&0\\
0&0&0&0&0&R_{0}&0&R_{0}\\
\end{array}\!\!\right),
\end{eqnarray}
\begin{eqnarray}\label{rho_TC}
\rho'^ {T_{C}}_{ABC}=
 \left(\!\!\begin{array}{cccccccc}
R_{0}&0&0&0&R_{0}&0&0&0\\
0&R_{2}&0&0&R_{0}&R_{2}&R_{0}&0\\
0&0&0&0&0&0&0&0\\
0&0&0&0&0&0&0&0\\
R_{0}&R_{0}&0&0&R_{0}+R_{1}&R_{0}&R_{1}&0\\
0&R_{2}&0&0&R_{0}&R_{0}+R_{2}&R_{0}&R_{0}\\
0&R_{0}&0&0&R_{1}&R_{0}&R_{1}&0\\
0&0&0&0&0&R_{0}&0&R_{0}\\
\end{array}\!\!\right).
\end{eqnarray}
Since the expressions for the negativities are relatively complicated, their explicit forms are omitted here.


\begin{thebibliography}{99}
\bibitem{Q1}
C. H. Bennett, Quantum information, Phys. Scr. {\bf 1998}, 210 (1998).


\bibitem{Q12}
S. Zhou, P. Zhang, and Z. Yu, Environment-induced transitions in many-body quantum teleportation, Phys. Rev. Research {\bf8}, L012007 (2026).

\bibitem{Q13}
Y. Guo, X. Chen, H. Peng, and Q. Tian, Average correlation as a resource of quantum teleportation, Phys. Rev. A {\bf112}, 052403 (2025).

\bibitem{Q14}
R. Horodecki, P. Horodecki, M. Horodecki, and K. Horodecki, Quantum entanglement, Rev. Mod. Phys. {\bf81}, 865 (2009).

\bibitem{Q15}
K. Lee, F. Turro, and X. Yao, Quantum computing for energy correlators, Phys. Rev. D {\bf111}, 054514 (2025).

\bibitem{Q16}
I. D. Smith, B. Klaver, H. P. Nautrup, W. Lechner, and H. J. Briegel, Minimally universal parity quantum computing, Phys. Rev. A {\bf112}, 032606 (2025).


\bibitem{Q17}
M. Bozzio, C. Cr\'{e}peau, P. Wallden, and P. Walther, Quantum cryptography beyond key distribution: Theory and experiment, Rev. Mod. Phys. {\bf97}, 045006 (2025).


\bibitem{Q19}
H. Yu, B. Crockett, N. Montaut, S. Sciara, M. Chemnitz, S. T. Chu, B. E. Little, D. J. Moss, Z. Wang et al., Exploiting Nonlocal Correlations for Dispersion-Resilient Quantum Communications, Phys. Rev. Lett. {\bf134}, 220801 (2025).

\bibitem{Q20}
Z. Wang and L. Jiang, Passive Environment-Assisted Quantum Communication with GKP States, Phys. Rev. X {\bf15}, 021003 (2025).

\bibitem{Q21}
A. Gandotra, Z. Wang, A. A. Clerk, and L. Jiang, Quantum communication over bandwidth-and-time-limited channels, Phys. Rev. A {\bf113}, 032616 (2026).

\bibitem{Q36}
L. Amico, R. Fazio, A. Osterloh,  V. Vedral, Entanglement in many-body systems, Rev. Mod. Phys. {\bf80}, 517 (2008).

\bibitem{Q39}
J. W. Pan, Z. B. Chen, C. Y. Lu, H. Weinfurter, A. Zeilinger, and M. \.{Z}ukowski, Multiphoton entanglement and interferometry, Rev. Mod. Phys. {\bf84}, 777 (2012).


\bibitem{Q38}
H. Cao, C. Radhakrishnan, M. Su, M. M. Ali, C. Zhang, Y. F. Huang, T. Byrnes, C. F. Li, and G. C. Guo,  Fragility of quantum correlations and coherence in a multipartite photonic system, Phys. Rev. A {\bf102}, 012403 (2020).


\bibitem{Q40}
M. Plesch and V. Bu\u{z}ek, Entangled graphs: Bipartite entanglement in multiqubit systems, Phys. Rev. A {\bf67}, 012322
(2003).


\bibitem{Q42}
Y. Wu, L. Tian, W. Yao, S. Shi, X. Liu, B. Lu, Y. Wang, Y. Zheng, Continuous variable quantum teleportation network with star topology, Appl. Phys. Lett. {\bf 124}, 114002 (2024).

\bibitem{Q44}
H. Zhang, S.  Yang,  K. He, Star network quantum steering, Quantum Inf. Process. {\bf22}, 286 (2023).

\bibitem{Q46}
A. K. Kashi, M. Kues, Chip-fiber-chip quantum teleportation in a star-topology
quantum network, Light sci. Appl. {\bf14}, 349 (2025).

\bibitem{SDF1}
W. Liu, C. Wen, J. Wang, Lorentz violation alleviates gravitationally induced entanglement degradation, J. High Energ. Phys. {\bf2025}, 184 (2025).

\bibitem{SDF2}
S. Sen, A. Mukherjee, and S. Gangopadhyay, Entanglement degradation as a tool to detect signatures of modified gravity, Phys. Rev. D {\bf109}, 046012 (2024).

\bibitem{SDF3}
I. Fuentes-Schuller and R. B. Mann, Alice falls into a black hole: Entanglement in non-inertial frames, Phys. Rev. Lett. {\bf95}, 120404 (2005).

\bibitem{SDF4}
Q. Pan and J. Jing, Hawking radiation, entanglement, and teleportation in the background of an asymptotically flat static black hole, Phys. Rev. D {\bf78}, 065015 (2008).

\bibitem{SDF5}
H. Wu and L. Chen, Orbital angular momentum entanglement in noninertial reference frame, Phys. Rev. D {\bf107}, 065006 (2023).

\bibitem{SDF6}
S. M. Wu, X. W. Fan, R. D. Wang, H. Y. Wu, X. L. Huang and H. S. Zeng, Does Hawking effect always degrade fidelity of quantum teleportation in Schwarzschild spacetime?, J. High Energ. Phys. {\bf2023}, 232 (2023).


\bibitem{SDF7}
E. Mart\'{\i}n-Mart\'{\i}nez, L. J. Garay and J. Le\'{o}n, Unveiling quantum entanglement degradation near a Schwarzschild black hole,  Phys. Rev. D {\bf82}, 064006 (2010).

\bibitem{SDF8}
P. M. Alsing, I. Fuentes-Schuller, R. B. Mann and T. E. Tessier, Entanglement of Dirac fields in non-inertial frames, Phys. Rev. A {\bf74}, 032326 (2006).

\bibitem{SDF9}
H. Du, R. B. Mann, Fisher information as a probe of spacetime structure: relativistic quantum metrology in (A)dS, J. High Energ. Phys. {\bf2021}, 112 (2021).


\bibitem{SDF10}
M. M. Du, H. W. Li, S. T. Shen, X. J. Yan, X. Y. Li, L. Zhou, W. Zhong and Y. B. Sheng, Maximal steered coherence in the background of Schwarzschild space-time, Eur. Phys. J. C {\bf84}, 450 (2024).

\bibitem{SDF11}
S. Elghaayda, A. Ali, S. Al-Kuwari and M. Mansour, Physically accessible and inaccessible quantum correlations of Dirac fields in Schwarzschild spacetime, Phys. Lett. A {\bf525}, 129915 (2024).


\bibitem{SDF12}
J. K. Basak, D. Giataganas, S. Mondal, and W. Y. Wen, Reflected entropy and Markov gap in noninertial frames, Phys. Rev. D {\bf108}, 125009 (2023).

\bibitem{SDF13}
Q. Xiao, Y. Chen, T. Liu, Effects of Lorentz symmetry breaking on quantum coherence in an expanding universe,
Phys. Lett. B {\bf872}, 140133  (2026).

\bibitem{SDF14}
C. Y. Liu, Z. W. Long and Q. L. He, Quantum coherence and quantum Fisher information of Dirac particles in curved spacetime under decoherence, Phys. Lett. B {\bf857}, 138991 (2024).


\bibitem{SDF15}
A. A. Svidzinsky, M. O. Scully, and W. Unruh, Minkowski vacuum entanglement and accelerated oscillator chains, Phys. Rev. D {\bf111}, 045022 (2025).


\bibitem{SDF16}
S. M. Wu, H. Y. Wu, Y. X. Wang, J. Wang, Gaussian tripartite steering in Schwarzschild black hole, Phys. Lett. B {\bf865},  139493 (2025).

\bibitem{SDF17}
G. Adesso, I. Fuentes-Schuller, M. Ericsson, Continuous variable entanglement sharing in non-inertial frames, Phys. Rev. A {\bf76}, 062112 (2007).


\bibitem{SDF18}
T. Gonzalez-Raya, S. Pirandola and M. Sanz, Satellite-based entanglement distribution and quantum teleportation with continuous variables, Commun. Phys. {\bf7},  126 (2024).

\bibitem{SDF19}
G. Ciliberto, S. Emig, N. Pavloff, and M. Isoard, Violation of Bell inequalities in an analog black hole, Phys. Rev. A {\bf109}, 063325 (2024).

\bibitem{SDF20}
S. M. Wu and H. S. Zeng, Genuine tripartite nonlocality and entanglement in curved spacetime, Eur. Phys. J. C {\bf82}, 4 (2022).

\bibitem{SDF21}
X. Liu, C. Zeng, J. Wang, Generation of quantum entanglement in superposed diamond spacetime, Eur. Phys. J. C {\bf85}, 539 (2025).

\bibitem{SDF22}
I. Agullo, A. Delhom, and \'A. Parra-L\'opez, Toward the observation of entangled pairs in BEC analog expanding universes, Phys. Rev. D \textbf{110}, 125023 (2024).


\bibitem{SDF23}
S. Bellucci, V. Kh. Kotanjyan, and A. A. Saharian, Fermionic condensate and the mean energy-momentum tensor
in the Fulling-Rindler vacuum, Phys. Rev. D {\bf108}, 085014 (2023).

\bibitem{SDF24}
T. Y. Wang and D. Wang, Entropic uncertainty relations in Schwarzschild space-time, Phys. Lett. B {\bf855}, 138876 (2024).

\bibitem{SDF25}
H. E. Camblong, A. Chakraborty, P. Lopez-Duque, and C. R. Ord\'{o}\~{n}ez, Entanglement degradation in causal diamonds,
Phys. Rev. D {\bf109}, 105003 (2024).

\bibitem{SDF26}
B. Yu, X. Y. Yang, X. Hu, Z. X. Jin, X. Huang, Nonlocal advantage of quantum imaginarity in Schwarzschild spacetime, arXiv:2604.03633.

\bibitem{SDF27}
A. Belfiglio, O. Luongo, S. Mancini, Quantum entanglement in cosmology, Phys. Rep. {\bf1146}, 1 (2025).


\bibitem{SDF28}
X. Liu, W. Liu, Z. Liu, J. Wang, Harvesting correlations from BTZ black hole coupled to a Lorentz-violating vector field, J. High Energ. Phys. {\bf2025}, 94 (2025).


\bibitem{SDF29}
Z. D. Wei, W. Han, Y. J. Zhang, Z. X. Man, Y. J. Xia, and H. Fan, Effect of the gravitational redshift on the precision of phase estimation, Phys. Rev. D \textbf{111},  026007 (2025).


\bibitem{SDF30}
W. Izquierdo,  J. Beltran,  E. Arias, Enhancement of harvesting vacuum entanglement in Cosmic String Spacetime, J. High Energ. Phys. {\bf2025}, 49 (2025).


\bibitem{SDF31}
Z. Liu, W. Liu, X. Liu, J. Wang, Wormhole-Induced correlation: A Link Between Two Universes, Phys. Lett. B {\bf879}, 140667 (2026).

\bibitem{SDF32}
S. M. Wu, C. X. Wang, D. D. Liu, X. L. Huang, H. S. Zeng, Would quantum coherence be increased by curvature effect in de Sitter space?, J. High Energ. Phys. {\bf2023}, 115  (2023).

\bibitem{SDF33}
A. Chakraborty,  L. Hackl, and M. Zych, Entanglement harvesting in quantum superposed spacetime, Phys. Rev. D {\bf111}, 104052 (2025).

\bibitem{SDF34}
Y. Tang, W. Liu, Z. Liu, J. Wang, Can the signatures of quantum superposition be detected
through correlation harvesting?, J. High Energ. Phys. {\bf2026}, 45 (2026).

\bibitem{SDF35}
Y. Tang, W. Liu, J. Wang, Observational signature of Lorentz violation in acceleration radiation, Eur.
Phys. J. C {\bf 85}, 1108 (2025).


\bibitem{SDF36}
Z. Liu, Y. Li, Z. Tian, J. Wang, Scrambling-Enhanced Quantum Battery Charging in Black Hole Analogues, Adv. Sci. {\bf 11}, e20281 (2026).

\bibitem{SDF37}
M. M. Du, H. W. Li, Z. Tao, S. T. Shen, X. J. Yan, X. Y. Li, W. Zhong, Y. B. Sheng and L. Zhou, Basis-independent quantum coherence and its distribution under relativistic motion, Eur. Phys. J. C {\bf84}, 838 (2024).

\bibitem{SDF38}
S. Barman, I. Chakraborty, and S. Mukherjee, Signatures of gravitational wave memory in the radiative process of entangled quantum probes, Phys. Rev. D {\bf111}, 025021 (2025).


\bibitem{SDF39}
J. Foo,  R. B. Mann, and M. Zych, Entanglement amplification between superposed detectors in flat and curved spacetimes,
Phys. Rev. D {\bf103}, 065013  (2021).


\bibitem{SDF40}
J. Foo, C. S. Arabaci, M. Zych, and R. B. Mann, Quantum superpositions of Minkowski spacetime, Phys. Rev. D {\bf107}, 045014 (2023).


\bibitem{SDF41}
X. Liu, Z. Tian, J. Wang, J. Jing, Protecting quantum coherence of two-level atoms from vacuum fluctuations of electromagnetic field, Ann. Phys.  {\bf366}, 102 (2016).


\bibitem{SDF42}
J. Foo, C. S. Arabaci, M. Zych, and R. B. Mann, Quantum Signatures of Black Hole Mass Superpositions, Phys. Rev. Lett. {\bf129}, 181301  (2022).


\bibitem{SDF43}
Q. Liu, T. Liu, C. Wen, and J. Wang, Optimal quantum strategy for locating Unruh channels, Phys. Rev. A {\bf110}, 022428 (2024).


\bibitem{SDF45}
A. Ali, S. Al-Kuwari, M. Ghominejad, M. T. Rahim, D. Wang, and S. Haddadi, Quantum characteristics near event horizons, Phys. Rev. D {\bf110}, 064001 (2024).




\bibitem{SDF46}
S. Harikrishnan, S. Jambulingam, P. P. Rohde, and C. Radhakrishnan, Accessible and inaccessible quantum coherence in relativistic quantum systems, Phys. Rev. A {\bf105}, 052403 (2022).

\bibitem{SDF47}
G. W. Mi, X. Huang, S. M. Fei, T. Zhang, Genuine four-partite Bell nonlocality in the curved spacetime, Eur. Phys. J. C {\bf85}, 354 (2025).


\bibitem{SDF48}
A. J. Torres-Arenas, Q. Dong, G. H. Sun, W. C. Qiang and S. H. Dong, Entanglement measures of W-state in noninertial frames, Phys. Lett. B {\bf789}, 93 (2019).


\bibitem{SDF49}
W. M. Li, S. M. Wu, Bosonic and fermionic coherence of N-partite states in the background of a dilaton black hole,  J. High Energ. Phys. {\bf2024}, 144 (2024).


\bibitem{SDF50}
G. W. Mi, X. Huang, S. M. Fei, T. Zhang, Quantumness near the Schwarzschild black hole based on W-state,
Ann. Phys.  {\bf537}, e00397 (2025).

\bibitem{SDF51}
S. M.  Wu, X. W. Teng, W. M. Li, Y. X. Wang, J. Lu, Nonseparability of multipartite systems in dilaton black hole, JCAP {\bf09}, 030  (2025).


\bibitem{SDF52}
S. M. Wu, X. W. Teng, J. X. Li, S. H. Li, T. H. Liu, J. C. Wang, Genuinely accessible and inaccessible
entanglement in Schwarzschild black hole, Phys. Lett. B {\bf848}, 138334 (2024).


\bibitem{SDF53}
S. H. Li, S. H. Shang, S. M. Wu, Does acceleration always degrade quantum entanglement for tetrapartite Unruh-DeWitt detectors?, J. High Energ. Phys. {\bf2025}, 214  (2025).

\bibitem{SDF54}
R. Hamzehofi, M. Ashrafpour, D. Afshar, Genuine entanglement and quantum coherence of a multipartite W state
in non-inertial frames, Eur. Phys. J. Plus  {\bf140}, 986 (2025).

\bibitem{SDF55}
T. Fan, C. Wen, J. Jing, J. Wang, Genuine tripartite entanglement and geometric quantum discord
in entangled three-body Unruh-DeWitt detector system, Front. Phys. {\bf19}, 54201 (2024).


\bibitem{SDF56}
B. S. DeWitt, Quantum Gravity: The New Synthesis, Cambridge University Press (1979).

\bibitem{SDF57}
N. D. Birrell, P. C. W. Davies, Quantum Fields in Curved Space, Cambridge Monographs on Mathematical
Physics, Cambridge University Press (1984).

\bibitem{SDF58}
L. C. C\'{e}leri, A. G. S. Landulfo, R. M. Serra, and G. E. A. Matsas, Sudden change in quantum and classical
correlations and the Unruh effect, Phys. Rev. A {\bf81}, 062130 (2010).

\bibitem{SDF59}
J. He, Z. Y. Ding, J. D. Shi, T. Wu, Multipartite quantum coherence and distribution under the Unruh effect, Ann. Phys. {\bf530}, 1800167 (2018).


\bibitem{SDF60}
Y. C. Ou and H. Fan, Monogamy inequality in terms of negativity for three-qubit states, Phys. Rev. A {\bf75}, 062308 (2007).

\bibitem{SDF61}
V. Coffman, J. Kundu, and W. K. Wootters,
Distributed entanglement, Phys. Rev. A {\bf61}, 052306
(2000).

\end{thebibliography}
\end{document}